\documentclass{article}
\usepackage{emulateapj,apjfonts,psfig, epsfig}

\def\myputfigure#1#2#3#4#5#6%
{\hskip0.03\textwidth\vskip#5pt
\makebox[0pt]{\hskip#2in
\includegraphics[width=#3\textwidth,angle=#6]{#1}}\vskip#4pt\hfill}

\makeatother

\def\fun#1#2{\lower3.6pt\vbox{\baselineskip0pt\lineskip.9pt
\ialign{$\mathsurround=0pt#1\hfil##\hfil$\crcr#2\crcr\sim\crcr}}}
\def\lap{\mathrel{\mathpalette\fun <}}
\def\gap{\mathrel{\mathpalette\fun >}}

\def\radius{{\cal R}}
\def\mass{{\cal M}}
\def\luminosity{{\cal L}}
\def\flux{{\cal F}}

\def\msun{{\mass_\odot}}
\def\rsun{{\radius_\odot}}
\def\lsun{{\luminosity_\odot}}

\def\beq{\begin{equation}}
\def\eeq{\end{equation}}

\def\mh{M_\bullet}

\lefthead{Stellar Disruption Rates}
\righthead{}

\begin{document}

\title{Revised Rates of Stellar Disruption in Galactic Nuclei}

\author{Jianxiang Wang and David Merritt}
\affil{Department of Physics and Astronomy, Rutgers University,
New Brunswick, NJ 08903}

\begin{abstract}
We compute rates of tidal disruption of stars by supermassive black holes
in galactic nuclei, using downwardly revised black hole masses
from the $\mh-\sigma$ relation.
In galaxies with steep nuclear density profiles, 
which dominate the overall event rate,
the disruption frequency varies inversely with 
assumed black hole mass.
We compute a total rate for non-dwarf galaxies of
$\sim 10^{-5}$ yr$^{-1}$ Mpc$^{-3}$, about
a factor ten higher than in earlier studies.
Disruption rates are predicted to be highest in nucleated 
dwarf galaxies, assuming that such galaxies contain
black holes.
Monitoring of a rich galaxy cluster for
a few years could rule out the existence of
intermediate mass black holes in dwarf galaxies.
\end{abstract}

\section{Introduction}

Stars that pass sufficiently close to a supermassive black hole
will be tidally disrupted
(\cite{Hills75,FR76,LO79}).
Disruption of solar-type stars occurs at a distance 
$r_t\approx \rsun(\mh/\msun)^{1/3}$, with $\mh$ the black hole mass; 
for $\mh\lap 10^8M_\odot$, the tidal radius lies beyond
the black hole's event horizon
and disruption results in an energetic flare as the
bound stellar debris falls back onto the black hole.
Emission from the debris is expected to peak in the soft X-ray or UV
domains, to have a maximum luminosity of $\sim 10^{44}$ erg s$^{-1}\approx
10^{11}\lsun$, and to decay on a time scale of weeks to months
(\cite{Rees88,EK89,Ulmer99,Kim99}).
Detection of flares would constitute robust proof of the
existence of supermassive black holes and could conceivably allow
constraints to be placed on black hole masses and spins
(\cite{Rees98}).

The {\it ROSAT} All-Sky Survey detected soft X-ray outbursts
from a number of galaxies with no previous history of
of Seyfert activity.
Roughly half a dozen of these events had the properties of a 
tidal disruption flare (Komossa 2002 and references therein),
and follow-up optical spectroscopy of the candidate
galaxies confirmed that at least two were subsequently inactive
(\cite{Gezari03}).

The mean event rate inferred from these outbursts is roughly
consistent with theoretical predictions (\cite{Donley02}).
Detailed calculations of the tidal disruption rate in samples
of nearby galaxies have been published by
Syer \& Ulmer (1999, hereafter SU)
and Magorrian \& Tremaine (1999, hereafter MT).
Both groups took black hole masses from the 
Magorrian et al. (1998) demographic study, 
which found a mean ratio of black hole mass to 
bulge mass of $\sim 0.006$.
Following the discovery of the $\mh-\sigma$ relation
(\cite{FM00,Gebhardt00}),
the mean value of $\mh/M_{bulge}$ was revised downward,
to $\sim 0.001$ (\cite{MF01a,KG01}).
The lower mean value of $\mh/M_{bulge}$
resolved two outstanding discrepancies:
the factor $\sim 10$ difference between black hole masses in quiescent
and active galaxies with similar luminosities (\cite{Wandel99});
and the higher apparent density of black holes in nearby 
galaxies compared with what is needed to explain the integrated 
light from quasars (\cite{Richstone98}).

Here we examine the consequences of downwardly revised
black hole masses for the rate of stellar tidal disruptions
in galactic nuclei.
Published scaling relations 
(e.g. Frank \& Rees 1976; Cohn \& Kulsrud 1978)
suggest that stellar consumption rates in galactic
nuclei should scale as $\sim\mh^n, 4/3\lap n\lap 9/4$
when the other properties (density,
velocity dispersion) of the host galaxy are fixed.
Hence one might naively expect the lower values of $\mh$
to imply lower rates of stellar disruption.
Instead we find the opposite: in most galaxies,
and in particular in the galaxies with steep nuclear
density profiles that dominate the overall event rate,
decreasing the assumed value of $\mh$ leads to {\it higher}
rates of loss cone feeding.
We estimate a total tidal disruption rate among
non-dwarf galaxies that is
about an order of magnitude higher than in studies
based on the Magorrian et al. black hole masses.

This paper is laid out as follows.
\S 2 describes the galaxy sample and \S 3 reviews the
steady-state loss cone theory from which event rates
are computed.
The theory is applied in \S4, with the counter-intuitive
result that lower values of $\mh$ imply greater feeding
rates in most galaxies.
This result is analyzed in more detail in \S5, where
it is shown to be a generic property of steep power-law
nuclei.
We derive an accurate, analytical expression for the tidal
disruption rate in singular isothermal sphere nuclei.
In \S6 we present the implications of our results for the
overall rate of tidal flaring and show that the predicted
rate would be so high in dwarf galaxies that the presence
of black holes in these systems could be ruled out by just
a few years' monitoring of a rich galaxy cluster like Virgo.
\S7 sums up.

\section{Galaxy Sample}

Our basic sample is the set of 61 elliptical galaxies whose
surface brightness profiles were studied by Faber et al. (1997).
These authors fit the luminosity data to the parametric model
\beq
I(\xi)=I_b2^{\frac{\beta-\Gamma}{\alpha}}\xi^{-\Gamma}\left(1+\xi^{\alpha}\right)^{-\frac{\beta-\Gamma}{\alpha}},\ \ \ \ \xi \equiv R/r_b
\label{eq:Nuker}
 \eeq
where $r_b$ is the ``break radius,'' $I_b=I(r_b)$,
and $\Gamma$ is the logarithmic slope
of the surface brightness profile at small radii.
(Note that we adopt the ``theorist's convention'' in which $\gamma$
refers to the logarithmic slope of the central {\it space} density profile.)
For $51$ of these galaxies, Faber et al. (1997) give values
for each of the five parameters
$\mu_b,\alpha$, $\beta$, $\Gamma$ and $\Upsilon_V$; the latter is the
visual mass-to-light ratio assuming $H_0=80 $ km s$^{-1}$ Mpc$^{-1}$
and $\mu_b$ is the surface brightness at $r_b$ 
in visual magnitudes arcsec$^{-2}$.
For these 51 galaxies we computed the mass density profile via
Abel's equation:
 \beq
 \rho(r)=\Upsilon_V j(r)=-\frac{\Upsilon_V}{\pi} \int_{r}^{\infty}\frac{dI}{dR}\frac{dR}{\sqrt{R^2-r^2}}
 \eeq
with $j(r)$ the luminosity density.
Below we follow the convention of referring
to galaxies with $\Gamma \lap 0.2$ as ``core'' galaxies
and those with $\Gamma \gap 0.2$ as ``power-law'' galaxies.
\footnote{In fact we retain the Faber et al. (1997) classifications
in Table 1, which are slightly different.}
We note that even most core galaxies exhibit an approximately
power-law dependence of space density on radius for small $r$,
$\rho\sim r^{-\gamma}$ (\cite{MF96,Gebhardt96}).
The weak power-law dependence $\rho$ on $r$ in the core galaxies
is not well reproduced by deprojection of the fitting function 
(\ref{eq:Nuker}), which is a possible source of systematic error
in what follows.
Our sample (Table 1) contains 28 power-law galaxies and 23 core galaxies.

The gravitational potential $\psi(r)\equiv -\Phi(r)$ was computed via
\begin{mathletters}
\begin{eqnarray}
\psi(r)&=& \psi_*(r) + {G\mh\over r}, \\
\psi_*(r) &=& {4\pi G\over r} \int_{r}^{\infty}\rho(r') {r'}^2 dr
 + 4\pi G \int_{r}^{\infty}\rho(r')r'dr' \\
&=&\frac{GM(r)}{r}-4\pi G \int_{r}^\infty \frac{\Upsilon_V}{\pi}\int_{r'}^{\infty}\frac{dI(R)}{dR}\frac{dR}{\sqrt{R^2-{r'}^2}}r'dr' \nonumber \\
& & \\
&=&\frac{GM(r)}{r} + 2^{2+\frac{\beta-\gamma}{\alpha}}G\Upsilon_V 
 r_b^{\beta} \int_{r}^{\infty} (\Gamma r_b^{\alpha}+\beta {r'}^{\alpha})
{r'}^{-\gamma-1} \nonumber \\
& & (r_b^{\alpha}+{r'}^{\alpha})^{(-1-\frac{\beta-\gamma}{\alpha})}
 \sqrt{{r'}^2-r^2}\ dr'.
\end{eqnarray}
\end{mathletters}
The distribution function $f$, defined as the number density
of stars in phase space, was computed via Eddington's formula:
\beq
 f(\varepsilon)=\frac{1}{\sqrt{8}\pi^2m_\star}\frac{d}{d\varepsilon}\int_{0}^{\varepsilon} 
\frac{d\rho}{d\psi}\frac{d\psi}{\sqrt{\varepsilon-\psi}}
\eeq
with $m_\star$ the stellar mass and $\varepsilon\equiv -E$.
We follow MT in assuming an isotropic
velocity distribution.
Galaxies with $\Gamma\lap 0.05$ were found to have $f<0$ when
$\mh>0$; this is a consequence of the fact that an isotropic $f$ can 
not reproduce a shallow density profile around a point mass.
The 10 galaxies with negative $f$'s are included at the end of Table 1 
and not discussed further here. 

We define the sphere of influence of the black hole 
to have radius $r_h$, where
\beq
M_\star(r_h) = 2\mh
\eeq
and $M_\star$ is the mass in stars within $r$.
This definition is equivalent to $r_h=G\mh/\sigma^2$ when
$\rho(r)\propto r^{-2}$.
We further define $\varepsilon_h\equiv\psi(r_h)$.

Of the 41 galaxies in our sample with non-negative $f$'s,
18 have black hole masses tabulated in Magorrian et al. (1998).
These masses are given in column 11 of Table 1.
For the remaining galaxies, we give in column 11 black holes
masses computed from
\beq
\mh=0.006M_{bulge}
\eeq 
the mean relation between bulge mass and black hole mass
found by Magorrian et al. (1998).

A second way to estimate black hole masses is via the $\mh-\sigma$
relation (Ferrarese \& Merritt 2000; Gebhardt et al. 2000).
We adopt the updated version from the review of Merritt \& Ferrarese (2001b):
\beq
\mh = 1.48 \times 10^8 \msun \left ({\sigma_c\over 200\ \mathrm{km}\ \mathrm{s}^{-1}}\right)^{4.65};
\label{eq:ms}
\eeq
the errors in the normalizing coefficient and exponent are
$\pm 0.24\times 10^8\msun$ and $\pm 0.48$ respectively.
Equation (\ref{eq:ms}) was determined from a fit to the small sample of
galaxies in which the black hole's sphere of influence is
clearly resolved.
The parameter $\sigma_c$ in equation (\ref{eq:ms}) is the velocity dispersion
measured in an aperture of size $r_e/8$ centered on the nucleus, 
with $r_e$ the effective radius (\cite{FM00}).
We computed $\sigma_c$ from published measurements of the central
velocity dispersion following the prescription in 
Ferrarese \& Merritt (2000). 
The $\mh-\sigma$ masses are listed in column 13 of Table 1.

As can be seen in Table 1, and discussed in detail elsewhere
(e.g. Ferrarese \& Merritt 2000; Merritt \& Ferrarese 2001b),
the Magorrian et al. masses are systematically high compared with
masses computed via the $\mh-\sigma$ relation.
It is this discrepancy which motivated the current study;
both published studies of stellar disruption rates 
in galactic nuclei (SU, MT)
were based on the Magorrian et al. masses.

\section{Loss Cone Theory}

Stars of mass $m_\star$ and radius $r_\star$ that come within a distance
\begin{equation}
r_t=(\eta ^2 \frac{M_{\bullet}}{m_{\star}})^{1/3}r_{\star}
\label{eq:rt}
\end{equation}
of the black hole will be tidally disrupted;
$\eta \approx 0.844$ for an $n=3$ polytrope.
Following MT,
we define the ``consumption rate'' $\dot N$ as the rate at
which stars come within $r_t$, even if $r_t$ falls below
the Schwarzschild radius $2G\mh/c^2$; the latter occurs when
$\mh\gap 10^8\msun$.
(The largest consumption rates occur in small dense galaxies for
which $r_t>r_s$.)
In a spherical galaxy, stellar orbits lie within the consumption
loss cone if their energy $\varepsilon$
and angular momentum per unit mass $J$ satisfy
\begin{equation}
J^2 \leq J^2_{lc}(\varepsilon)\equiv 2r^2_t[\psi(r_t)-\varepsilon] \simeq 2GM_{\bullet} r_t.
\end{equation}

We adopt the Cohn-Kulsrud (1978; hereafter CK) 
formalism for computing the flux
of stars into the loss cone.
Let ${\cal F}(\varepsilon)$ be the number of stars per unit time 
and unit energy that are deflected into the loss cone via gravitational
encounters with other stars.
Define $\langle\left (\Delta R\right)^2\rangle$ to be the diffusion
coefficient in $R\equiv J^2/J_c^2(\varepsilon)$, with $J_c(\varepsilon)$
the angular momentum of a circular orbit of energy $\varepsilon$.
Then (CK)
\beq
{\cal F}(\varepsilon)d\varepsilon = 4\pi^2 J_c^2  
\left\{
\oint \frac{dr}{v_r}
\lim_{R\rightarrow 0}
\frac{\langle(\Delta R)^2\rangle}{2R} 
\right\}
\frac{f}{\ln R_0^{-1}}d\varepsilon
\label{eq:CK}
\eeq
and the total consumption rate is given by 
\begin{equation}
\dot N=\int {\cal F}(\varepsilon)d\varepsilon.
\end{equation}
In equation (\ref{eq:CK}),
$R_0(\varepsilon)$ is the value of $R$ at which $f$ falls to
zero due to removal of stars that scatter into the loss cone.
$R_0$ is not equal to its geometrical value,
$R_{lc} = J_{lc}^2/J_c(\varepsilon)^2$,
because scattering of stars {\it onto} loss cone orbits permits
$f$ to be nonzero even for $J<J_{lc}$.
CK find that $R_0(\varepsilon)$ can be well approximated
by
\beq
R_0(\varepsilon) = R_{lc}(\varepsilon) \times 
\cases{ \exp(-q), & $q(\varepsilon) > 1$ \cr
  \exp(-0.186 q -0.824 \sqrt{q}), & $q(\varepsilon) < 1$} ,
\label{eq_R0}
\eeq
with 
\beq
q(\varepsilon)\equiv \frac{1}{R_{lc}(\varepsilon)} \oint 
\frac{dr}{v_r}
\lim_{R\rightarrow 0}
\frac{\langle(\Delta R)^2\rangle}{2R} = {P(\varepsilon) \overline{\mu}(\varepsilon)\over R_{lc}(\varepsilon)};
\eeq
$P(\varepsilon)$ is the period of a radial orbit with energy $\varepsilon$
and $\overline{\mu}$ is the orbit-averaged
diffusion coefficient.
The function $q(\varepsilon)$ can be interpreted as the ratio
of the orbital period to the time scale for diffusional refilling
of the loss cone; $q\gg 1$ defines the ``pinhole'' or ``full loss cone''
regime in which encounters replenish loss cone orbits much more rapidly
than they are depleted.
MT give expressions for the local angular-momentum
diffusion coefficient:
\begin{mathletters}
\begin{eqnarray}
&&\lim_{R\rightarrow 0}{\langle\left(\Delta R\right)^2\rangle\over 2R} = 
{32\pi^2r^2G^2m_\star^2\ln\Lambda\over 3J_c^2}\left(3I_{1/2}-I_{3/2}+2I_0\right), \nonumber \\
& & \\
I_0 & \equiv& \int_0^\varepsilon f(\varepsilon')d\varepsilon', \\
I_{n\over 2} &\equiv&\left[2\left(\psi(r)-\varepsilon\right)\right]^{-{n\over 2}} \int_\varepsilon^{\psi(r)}\left[2\left(\psi(r)-\varepsilon'\right)\right]^{3/2}f(\varepsilon')d\varepsilon'
\end{eqnarray}
\end{mathletters}
from which the orbit-averaged quanties 
$\overline\mu(\varepsilon)$ and $q(\varepsilon)$ can be computed.

In the Fokker-Planck approximation under which equation (\ref{eq:CK})
was derived,
the flux of stars into the loss cone at each energy is determined by
gradients in $f$ with respect to $R$ at $R\approx R_{lc}$.
CK, who modelled globular clusters,
derived expressions for these gradients by assuming that the distribution
of stars near the loss cone had evolved to a steady state
in which the encounter-driven supply of stars into the loss cone
was balanced by consumption.
Relaxation times in galactic nuclei are usually in excess of a Hubble time,
particularly in the core galaxies (\cite{Faber97}),
and it is not clear that the distribution of stars near the loss cone
will have had time to reach a steady state in all of our galaxies
(\cite{MM03}).
We will return to this question in a subsequent paper; for
now, we follow MT in
assuming that the CK loss cone boundary solution applies
to galactic nuclei.

\section{Dependence of the Consumption Rate on $\mh$ in the Galaxy Sample}

Figures 1 and 2 show how the energy dependence of various quantities
changes with the assumed value of $\mh$ in two galaxies:
NGC 4551, a power-law galaxy ($\Gamma = 0.8$); and
NGC 4168, a core galaxy ($\Gamma = 0.14$).
The value of $\Upsilon_V$, and hence the mass density of the stars,
was fixed as $\mh$ was varied.
We adopted $m_\star=\msun$ throughout.
In the power-law galaxy, as $\mh$ is decreased,
the flux of stars into the loss cone increases at 
$\varepsilon\gap\varepsilon_h$ and decreases at
$\varepsilon\lap\varepsilon_h$; since most of the
flux comes from near the black hole,
$\varepsilon\gap\varepsilon_h$, the total consumption rate
increases with decreasing $\mh$.
In the core galaxy, the dependence of $\flux(\varepsilon)$ on
$\mh$ is more complex: $\flux(\varepsilon\approx\varepsilon_h)$
first increases, then decreases, with decreasing $\mh$.
The consequences of these trends can be seen in Figure 3, 
which plots integrated 
consumption rates $\dot N$ 
\myputfigure{f1.ps}{2.75}{0.43}{-10}{-10}{0.}
\figcaption{\label{fig:4551}
Dependence on $\mh$ of various quantities associated with
stellar consumption in the power-law galaxy NGC 4551. 
Stars are assumed to have solar mass and radius.
Left column: $r_{apo}$ (apocenter radius of radial orbit);
$f$ (phase space number density); $q$ (quantity that distinguishes
between diffusion and full-loss-cone regimes).
Right column: $P$ (period of radial orbit);
$R_{lc}$ (geometric size of loss cone in terms of $R\equiv J^2/J_c^2$);
$\flux$ (flux into loss cone).
As $\mh$ is reduced, more and more of the galaxy falls within
the full-loss-cone regime ($q\gg 1$) and the total flux of stars
into the loss cone rises.
}
\vspace{\baselineskip}
\myputfigure{f2.ps}{2.75}{0.43}{-10}{-10}{0.}
\figcaption{\label{fig:4168}
Like Figure \ref{fig:4551}, but for the core galaxy NGC 4168.
By comparison with NGC 4551,
less of the galaxy lies in the full-loss-cone regime 
and the total consumption rate is lower.
}
\vspace{\baselineskip}

\noindent
as a function of $\mh$ for every galaxy 
in our sample.
The power-law galaxies exhibit monotonic or nearly montonic
trends of
increasing $\dot N$ with decreasing $\mh$;
the dependence is roughly $\dot N\propto\mh^{-1}$.
In the case of the core galaxies, $\dot N$ generally increases
with $\mh$ up to a maximum value, 
then decreases as $\mh$ is increased further.
This behavior is explained in the
next section.

\myputfigure{f3.ps}{3.}{0.4}{0}{0}{-90.}
\figcaption{\label{fig:allflux}
Dependence of consumption rate on assumed black hole mass for the
galaxies in Table 1.}
\vspace{\baselineskip}

Our primary concern is how consumption rates
would change if the black hole masses used by earlier
authors were replaced with the presumably more
accurate masses derived from the $\mh-\sigma$ relation.
Figure 4 makes this comparison.
In almost every galaxy in our sample, 
the inferred $\dot N$ is greater
when the $\mh-\sigma$ black hole mass is used.
The changes are greatest in the power-law galaxies,
since $\dot N\sim\mh^{-1}$ in these galaxies.
In the core galaxies, although $\mh$ is sometimes
increased by as much as $10^2$, the changes in $\dot N$
are usually modest because of the nearly flat dependence
of $\dot N$ on $\mh$ in these galaxies (Figure 3).

\myputfigure{f4.ps}{3.}{0.35}{0}{0}{-90.}
\figcaption{\label{fig:masscp}
Comparison of consumption rates computed using the two values
of $\mh$ in Table 1.
Abscissa: $\mh$ computed from the $\mh-\sigma$ relation, 
equation (\ref{eq:ms}).
Ordinate: $\mh$ from Magorrian et al. (1998).
}
\vspace{\baselineskip}

Figure \ref{fig:mlcurrent} shows the dependence of 
$\dot N$ on galaxy luminosity and $\mh$ in our sample;
black hole masses were taken from the $\mh-\sigma$ relation.
The dependence of flaring rate on $\mh$ is fairly tight,
with a mean slope of $\dot N\sim\mh^{-0.8}$.

\myputfigure{f5.ps}{3.}{0.5}{10}{0}{0.}
\figcaption{\label{fig:mlcurrent}
Consumption rate as a function of galaxy luminosity (a) and black 
hole mass (b).
The dashed line in (b) is the relation defined by the singular
isothermal sphere, eq. (29);
it is a good fit to the galaxies plotted with stars,
which have steep central density profiles,
$\rho\sim r^{-2}$.
}
\vspace{\baselineskip}

Figure \ref{fig:analytic} shows how $\dot N$ and three critical radii 
associated with the black hole vary with $\mh$ in NGC 4551 and NGC 4168.
The tidal radius $r_t$ and the radius of influence $r_h$
were defined above.
The third radius, $r_{crit}$, is defined as:
\begin{equation}
\psi(r_{crit})=\varepsilon_{crit},\ \ \ \ q(\varepsilon_{crit})=1.
\end{equation}
$r_{crit}$ is roughly the radius of transition between the
``diffusive'' ($q<1$) and ``full-loss-cone'' ($q>1$) regimes,
and most of the flux into the loss cone
comes from radii $r\lap r_{crit}$.
We note that $r_{crit}\lap r_h$ over the relevant range in $\mh$
for both galaxies, and that both $r_t$ and $r_{crit}$ are less than $r_b$.
These same inequalities were found to hold for most of the galaxies
in our sample, which motivated the simplified treatment in the following 
section.

\section {Dependence of Consumption Rate on $\mh$ in Power-Law Nuclei}

We noted above the curious behavior of $\dot N$ as $\mh$ is varied
in a galaxy with otherwise fixed properties: $\dot N$ generally
increases as $\mh$ is reduced.
Here we show how the dependence of $\dot N$ on $\mh$ 
can be understood.
We derive the exact consumption rate in a $\rho\propto r^{-2}$ galaxy,
which is a good model for the faintest galaxies in our sample,
then derive approximate scaling relations for the dependence of 
$\dot N$ on $\mh$ in nuclei with shallower power-law indices.

\myputfigure{f6.ps}{3.}{0.5}{-10}{-10}{0.}
\figcaption{\label{fig:analytic}
Top panels: dependence of consumption rate 
on $\mh$ in NGC 4551 (power-law galaxy) and
NGC 4168 (core galaxy).
Dashed lines are the approximate $\mh$-dependences derived
in \S 5.2.
Bottom panels: $\mh$-dependence of three characteristic
radii: $r_t$, the tidal disruption radius; 
$r_{crit}$, the radius dividing the diffusive- and
full-loss-cone regimes; 
and $r_h$, the black hole's radius of influence.
$r_b$ is the break radius of the luminosity profile.
}
\vspace{\baselineskip}

\subsection{The Singular Isothermal Sphere}

Faint galaxies like M32 have the greatest consumption rates.
These galaxies also have steep power-law nuclear density
profiles, $\rho\sim r^{-\gamma}$, $\gamma\approx 2$.
Since $r_{crit}$ is generally less than $r_b$ in our galaxies
(cf. Figure \ref{fig:analytic}), 
we can approximate the stellar density profile as a single power law
when computing the flux.
Here we consider the singular isothermal sphere (SIS), $\gamma=2$.
The SIS density profile and potential are
\beq
\rho(r) = {\sigma^2\over 2\pi Gr^2},\ \ \ \ \psi_*(r) = -2\sigma^2\ln\left({r\over r_h}\right), \ \ \ \ r_h \equiv {G\mh\over\sigma^2}
\label{eq:SIS}
\eeq
where $\sigma$ is the 1D stellar velocity dispersion, independent
of radius for $r\gap r_h$.
The potential due to the stars has been normalized to zero
at $r=r_h$, and $\psi(r) = \psi_*(r) + G\mh/r$.
The isotropic distribution function describing the stars is
\begin{mathletters}
\begin{eqnarray}
f(\varepsilon)&=&\frac{1}{\sqrt{8}\pi^2m_{\star}}\frac{d}{d\varepsilon}\int_{0}^{\varepsilon} 
 \frac{d\rho}{d\psi}\frac{d\psi}{\sqrt{\varepsilon-\psi}} \\
&=& {1\over r_h^3\sigma^3} \left({\mh\over m_\star}\right) g^*(\varepsilon^*), \\
g^*(\varepsilon^*) &=& {\sqrt{2}\over 4\pi^3}\int_{-\infty}^{\varepsilon^*} {L^2(u)\left[2+L(u)\right]\over \left[1+L(u)\right]^3} {d\psi^*\over\sqrt{\varepsilon^* - \psi^*}}, \nonumber \\ 
u(\psi^*) &\equiv& {1\over 2} e^{\psi^*/2}.
\label{eq:gofe}
\end{eqnarray}
\end{mathletters}
\noindent
$L(u)$ is the Lambert function (also called the $W$ function) defined 
implicitly via $u=Le^L$.
The superscript ``$*$'' denotes dimensionless quantities
and the units of mass and velocity are
\beq
\left[M\right] = \mh,\ \ \ \ \left[V\right] = \sigma
\eeq
with $G=1$.
The dimensionless function $q(\varepsilon^*)$ that characterizes the 
deflection amplitude per orbital period is
\beq
q(\varepsilon^*)={32\pi^2\over 3\sqrt{2}}\ln\Lambda\left({m_\star\over\mh}\right) {h^*(\varepsilon^*)\over \psi^*(r_t) - \varepsilon^*} \left({r_t\over r_h}\right)^{-2}
\label{eq:qofe}
\eeq
and the loss cone flux is
\beq
\flux^*(\varepsilon^*) = {256\pi^4\over 3\sqrt{2}} {\ln\Lambda\over\ln R_0^{-1}} g^*(\varepsilon^*) h^*(\varepsilon^*),
\label{eq:flux}
\eeq
where
\begin{mathletters}
\begin{eqnarray}
h^*(\varepsilon^*) &=& h_1^*(\varepsilon^*) + h_2^*(\varepsilon^*) + h_3^*(\varepsilon^*), \\
h_1^*(\varepsilon^*) &=& 2\left[\int_{-\infty}^{\varepsilon^*} g^*({\varepsilon^*}') d{\varepsilon^*}'\right] \left[\int_0^{r^*({\varepsilon^*}')} {dr^* {r^*}^2\over\sqrt{\psi^*(r^*) - \varepsilon^*}}\right],
\\
h_2^*(\varepsilon^*) &=& 3\int_0^{r^*(\varepsilon^*)} {dr^* {r^*}^2\over\psi^*(r^*) - \varepsilon^*} \int_{\varepsilon^*}^{\psi^*(r^*)} d{\varepsilon^*}' \sqrt{\psi^*(r^*) - {\varepsilon^*}'} g^*({\varepsilon^*}'),
\\
h_3^*(\varepsilon^*) &=& -\int_0^{r^*(\varepsilon^*)} {dr^* {r^*}^2\over\left[\psi^*(r^*) - \varepsilon^*\right]^2} \int_{\varepsilon^*}^{\psi^*(r^*)} d{\varepsilon^*}' \left[\psi^*(r^*) - {\varepsilon^*}'\right]^{3/2} g^*({\varepsilon^*}').
\end{eqnarray}
\end{mathletters}
\noindent

\myputfigure{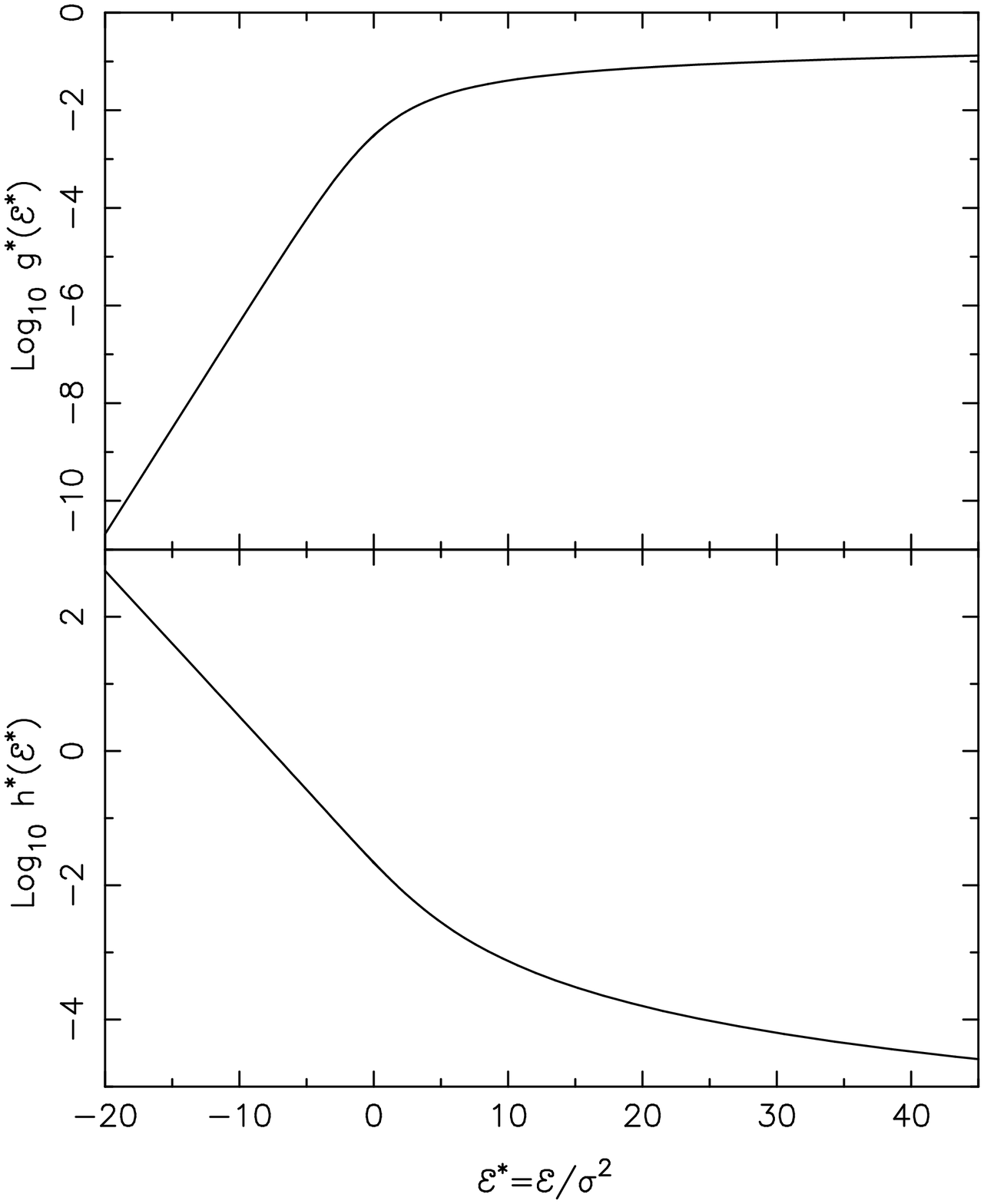}{3.}{0.3}{-10}{-10}{0.}
\figcaption{\label{fig:ghe}
Dimensionless functions $g^*(\varepsilon^*)$, $h^*(\varepsilon^*)$ 
that characterize the phase-space density and angular momentum
diffusion coefficient (equations \ref{eq:gofe},21)
in a singular isothermal sphere galaxy.}
\vspace{\baselineskip}

\noindent
Note that the functions $g^*(\varepsilon^*)$ and $h^*(\varepsilon^*)$ are
determined uniquely in these dimensionless units.
These functions are plotted in Figure \ref{fig:ghe}.

\myputfigure{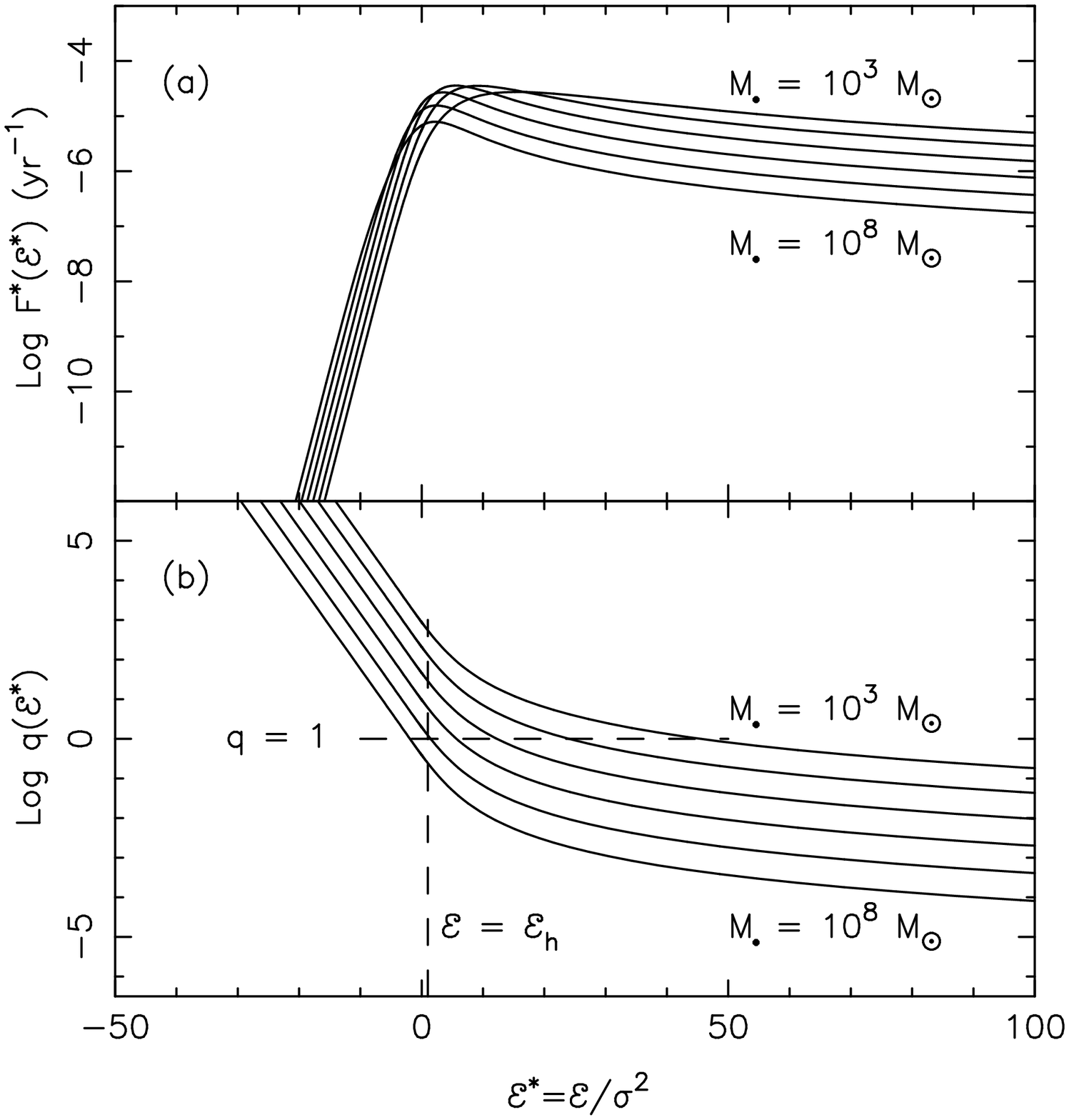}{3.}{0.4}{-10}{-10}{0.}
\figcaption{\label{fig:flux1}
Dimensionless loss-cone flux $\flux^*$ (equation \ref{eq:flux})
and $q$ (equation \ref{eq:qofe})
as functions of energy in a singular isothermal sphere galaxy, 
for various values of $\mh$.
The $\mh-\sigma$ relation (equation \ref{eq:ms})
was used to relate $\sigma$ to $\mh$.
}
\vspace{\baselineskip}

The function $R_0(\varepsilon)$ that defines the
edge of the loss cone is given by
equation (\ref{eq_R0}), with
\beq
R_{lc}(\varepsilon) = 2\left({r_t\over r_h}\right)^2 {\psi^*(r_t^*)-\varepsilon^*\over \left(2+{r_c^*}^{-1}\right) {r_c^*}^2},\ \ \ \ 
r_c^*(\varepsilon^*)=1/4L\left[e^{-(1-\varepsilon^*)/2}/4\right];
\eeq
$r_c(\varepsilon)$ is the radius of a circular orbit of energy $\varepsilon$.

If we set $\Lambda = 0.4\mh/m_*$ (\cite{SH71}),
the dimensionless flux $\flux^*(\varepsilon^*)$ is determined by the two 
parameters
\beq
\left({\mh\over m_*},\ \ {r_h\over r_t}\right).
\eeq
Adopting equation (\ref{eq:rt}) for $r_t$, we can write the second of these
two parameters as
\begin{mathletters}
\begin{eqnarray}
{r_h\over r_t} &=& {2\Theta\over\eta^{2/3}}\left({\mh\over m_*}\right)^{2/3} \\
&=& 21.5 \left({\mh\over m_*}\right)^{2/3}\left({\sigma\over 100\ \mathrm{km\ s}^{-1}}\right)^{-2} \left({m_\star\over\msun}\right) \left({r_\star\over\rsun}\right)^{-1}
\label{eq:rbrt}
\end{eqnarray}
\end{mathletters}
\noindent
with $\Theta\equiv Gm_\star/2\sigma^2 r_*$ the Safronov number; 
$\eta$ has been set to $0.844$.
Given values for $m_\star$ and $r_\star$, the two parameters
that specify $\flux(E)$ are then $\mh$ and $\sigma$.

Figure \ref{fig:flux1}a shows $\flux^*(\varepsilon^*)$ for various values of
$\mh$; $\sigma$ was computed from $\mh$ via the $\mh-\sigma$ relation
(\ref{eq:ms}).
The flux exhibits a mild maximum at $\varepsilon\approx \varepsilon_h$ and 
falls off slowly toward large (bound) energies.
The function $q(\varepsilon^*)$ is shown in Figure \ref{fig:flux1}b.
As $\mh$ is reduced, more and more of the nucleus lies within the
full-loss-cone regime, $q\gg 1$.

\myputfigure{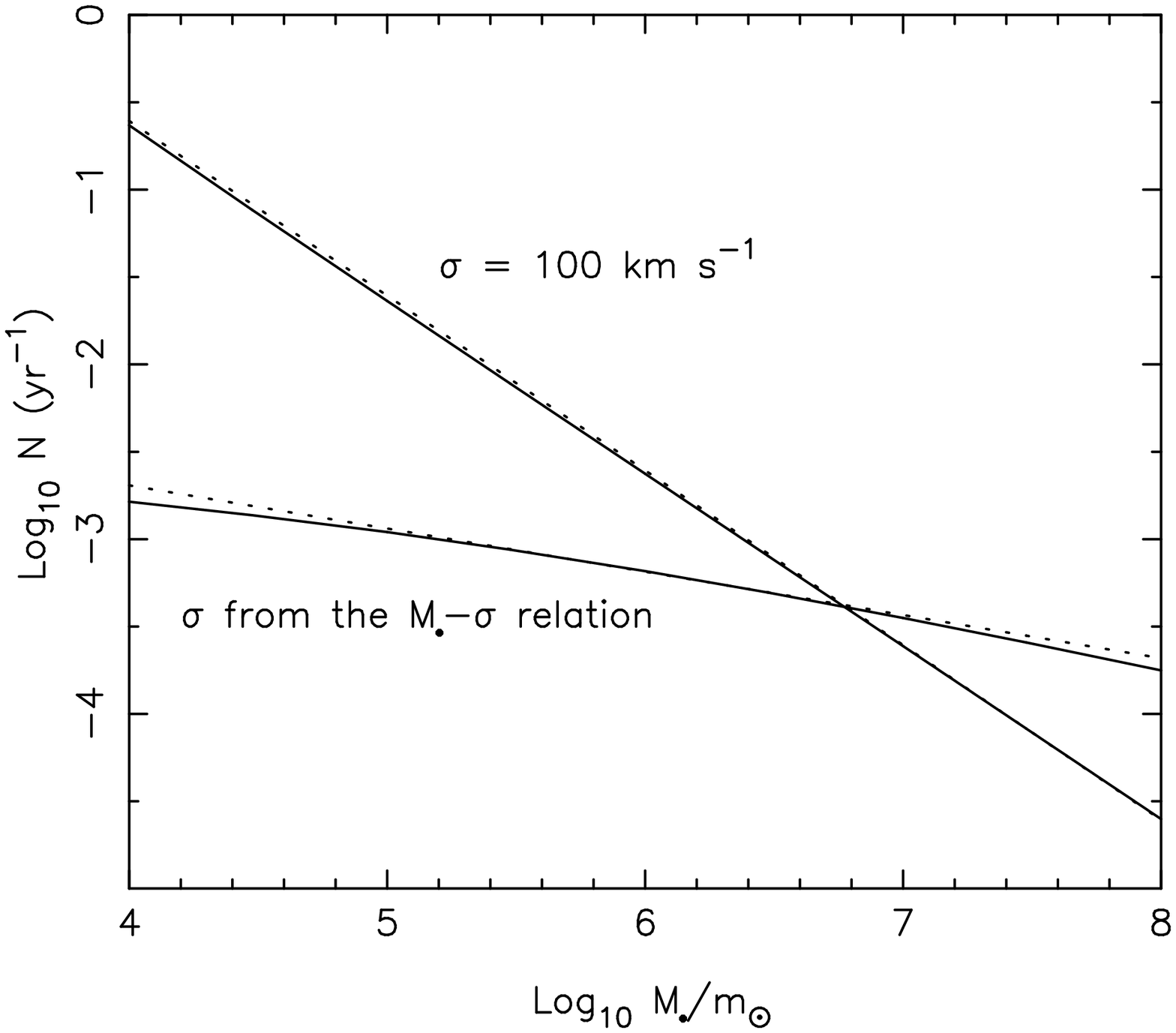}{3.}{0.5}{-80}{-10}{0.}
\figcaption{\label{fig:flux2}
Consumption rate as a function of $\mh$ in singular isothermal sphere
nuclei,
under two assumptions about $\sigma(\mh)$.
Dashed lines: equation (29).
}
\vspace{\baselineskip}

Figure \ref{fig:flux2} shows the consumption rate $\dot N = \int\flux(E) dE$
as a function of $\mh$
under two assumptions about the relation of $\sigma$ to $\mh$:
$\sigma=100$ km s$^{-1}$; 
or $\sigma$ determined from the $\mh-\sigma$ relation (\ref{eq:ms}).
For fixed $\sigma$, Figure \ref{fig:flux2} shows that 
$\dot N\sim \mh^{-1}$,
while allowing $\sigma$ to vary with $\mh$ implies a weaker 
(but still inverse) dependence of $\dot N$ on $\mh$.

The scaling of the consumption rate with $\mh$ and $\sigma$ can be
derived in a straightforward way.
Figure \ref{fig:flux1} shows that over a wide range of $\mh$ values,
most of the flux comes from $\varepsilon\gap \varepsilon_h$.
In this energy interval, $\psi^*(r^*) \approx G\mh/r^*$ and
\beq
g^*(\varepsilon^*) \approx {1\over\sqrt{2}\pi^3} \varepsilon^{1/2},
\ \ \ \ \ \ 
h^*(\varepsilon^*) \approx {5\sqrt{2}\over 24\pi^2} {\varepsilon^*}^{-2};
\eeq
the latter expression makes use of the fact that $h^*\approx h_1^*$,
i.e. most of the flux comes from scattering by stars with energies
greater than that of the test star.
Thus
\beq
q(\varepsilon^*) \approx {20\over 9} \ln\Lambda \left({m_\star\over\mh}\right) \left({r_h\over r_t}\right) {\varepsilon^*}^{-2}
\eeq
and $R_{lc}\approx 4(r_t/r_h)\varepsilon^*$.
The dimensionless flux is
\begin{mathletters}
\begin{eqnarray}
\flux^*(\varepsilon^*) &\approx& {160\ln\Lambda\over 9\sqrt{2}\pi} {\varepsilon^*}^{1/2} \left[A + {\varepsilon^*}^2\ln\left({B\over\varepsilon^*}\right)\right]^{-1},\label{eq:app1a} \\
A&\equiv& {20\over 9}\ln\Lambda \left({m_\star\over\mh}\right) \left({r_h\over r_t}\right),\\
B&\equiv& {r_h\over 4r_t}.
\end{eqnarray}
\end{mathletters}
\noindent
Ignoring the weak $E$-dependence of the logarithmic term and taking
$r_h/r_t$ from equation (\ref{eq:rbrt}), we find
\beq
\dot N=\int\flux(E)dE\propto A^{-1/4}\propto \sigma^{7/2}\mh^{-11/12}.
\label{eq:ndotapp}
\eeq
After some experimentation, we found that the following, slightly
different scaling:
\beq
\dot N \approx 7.1\times 10^{-4} {\rm yr}^{-1} 
\left({\sigma\over 70\ {\rm km\ s}^{-1}}\right)^{7/2} 
\left({\mh\over 10^6\msun}\right)^{-1},
\label{eq:fapp}
\eeq
provides a better fit to the exact, numerically-computed feeding rates
over the relevant range in $\mh$ (Figure \ref{fig:flux2}).
The normalization constant in equation (\ref{eq:fapp}) was chosen
to reproduce $\dot N$ exactly for $\sigma=70$ km s$^{-1}$, 
$\mh=10^6\msun$.
The consumption rate in equation (\ref{eq:fapp}) was derived
assuming stars of solar mass and radius and
scales as $m_\star^{-1/3} r_\star^{1/4}$.

\subsection{Shallower Power-Law Profiles}

The bright galaxies in our sample have nuclear density profiles
with shallower power-law indices, $\rho\sim r^{-\gamma}$,
$\gamma\lap 2$.
We calculate the
dependence of $\dot N$ on $\mh$ in these galaxies
using a more approximate approach.

The flux of stars into the loss cone, equation (\ref{eq:CK}), 
can be written
\begin{mathletters}
\begin{eqnarray}
{\cal F}(E) &=& {{\cal F}_{max}(E)\over\ln R_0^{-1}}, \\
{\cal F}_{max}(E) &\equiv& 4\pi^2 P(E)J_c^2(E) \overline{\mu}(E) f(E) \\
&\approx& \overline{\mu}(E) N(E)
\label{eq:32c}
\end{eqnarray}
\end{mathletters}
\noindent
with $N(E)$ the number of stars per unit energy interval;
equation (\ref{eq:32c}) assumes that $P(E,J)\approx P(E)$.
Now $\mu \equiv 2 r^2\langle\Delta v_t^2\rangle/ J_c^2$,
a function both of $r$ and $E$, 
and its orbit-averaged value $\overline{\mu}$ can be interpreted
as the time-averaged inverse of the relaxation time $T_R$ for 
orbits of energy $E$.
Hence
\beq
{\cal F}(E) \approx {N(E)\over T_R(E)} {1\over \ln R_0^{-1}(E)}.
\eeq
Above some energy $E_{crit}$,
$R_0\approx R_{lc} e^{-q}$ falls off rapidly with increasing $E$,
while for $E\lap E_{crit}$, $R_0\approx R_{lc}$ and 
$\ln R_0^{-1}$ is a slowly-varying function of order unity.
Hence
\begin{mathletters}
\begin{eqnarray}
\dot N &\approx& \int^{E_{crit}}_{-\infty} {N(E)\over T_R(E)}dE
\label{eq:FR1} \\
&\approx& {N(r<r_{crit})\over T_R(r_{crit})}
\label{eq:FR2}
\end{eqnarray}
\end{mathletters}
where $\Phi(r_{crit})\approx E_{crit}$.
These expressions correspond physically to the fact that the time
to scatter into the loss cone is comparable to $T_R$
for all $E\lap E_{crit}$.
Frank \& Rees (1977, hereafter FR) used an equation similar to
(\ref{eq:FR2})
to estimate feeding rates in nuclei with constant-density cores.
We repeat their analysis here, for black holes in nuclei with
arbitrary density slopes:
\beq
\rho(r) = \rho_0\left({r\over r_0}\right)^{-\gamma}.
\eeq

For $T_R$ we take the Spitzer \& Harm (1958) reference time:
\beq
T_R(r)= {\sqrt{2}\sigma^3(r) \over \pi G^2 m_\star\rho(r)\ln\Lambda}.
\eeq
Since $r_{crit}\lap r_h$ (Figure \ref{fig:analytic}), 
we can write $\sigma^2(r)\approx G\mh/r$ and
\beq
\dot N\approx {N(r<r_{crit})\over T_R(r_{crit})} \propto 
(3-\gamma)^{-1}\ln\Lambda G^{1/2}\rho_0^2r_0^{9/2}\mh^{-3/2}\left({r_{crit}\over r_0}\right)^{9/2-2\gamma}.
\eeq
Following FR,
we define $r_{crit}$ to be the radius above which encounters can 
scatter stars into or out
of the loss cone in a single orbital period;
at this radius, $q\approx 1$.
At $r=r_{crit}$, the angular size of the loss cone, $\theta_{lc}$,
is comparable to the angle $\theta_d$ by which a star is deflected in a
single period;
taking account of gravitational focussing, $\theta_{lc} \approx 
\sqrt{r_t/r}$.
We adopt equation (\ref{eq:rt}) for $r_t$ and write
$\theta_d^2\approx P/T_r \approx \sqrt{r^3/G\mh}/T_r$; 
the square root dependence of $\theta_d$
on $P$ reflects the fact that entry into the 
loss cone is a diffusive process.
Setting $\theta_{lc}=\theta_d$ then gives
\beq
\left({r_{crit}\over r_0}\right)^{4-\gamma} = 
{\rm Constant}\times \left(\ln\Lambda\right)^{-1} m_\star^{-1}\rho_0^{-1}r_0^{-4}\mh^{7/3}
\eeq
and for fixed $(\rho_0,r_0)$,
\beq
\dot N \propto \mh^\delta,\ \ \ \ \delta = {27-19\gamma\over 6(4-\gamma)}.
\label{eq:ndot}
\eeq
For the SIS, $\gamma=2$ and we recover $\delta = -11/12$ 
(equation \ref{eq:ndotapp}).

For $\gamma<27/19=1.42$, equation (\ref{eq:ndot}) gives $\delta>0$
and $\dot N$ increases with increasing $\mh$; for instance,
setting $\gamma=0$ gives the constant-density core and
$\dot N\propto \mh^{1.1}$.
This explains why the tidal destruction rates in the ``core'' galaxies
generally increase with increasing $\mh$.
As $\mh$ is increased still further in these galaxies,
$\dot N$ drops, since $r_h>r_b$ and the effective power-law 
index becomes steeper.
Figure 6 shows fits of equation (\ref{eq:ndotapp}) to $\dot N(\mh)$
for two galaxies.

\section{Implications for the Detection of Flares}

Black hole mass is observed to correlate tightly with bulge 
velocity dispersion (\cite{FM00,Gebhardt00}),
and with bulge mass
(\cite{MF01a}) and luminosity (\cite{MD02,Erwin02}).
Hence we can convert our scaling relation (\ref{eq:fapp})
into a net scaling of $\dot N$ on $L$.
First combining the $\mh-\sigma$ relation (\ref{eq:ms})
with equation (\ref{eq:fapp}):
\begin{mathletters}
\begin{eqnarray}
\dot N &\approx& 4.2 \times 10^{-4} {\rm yr}^{-1} \left({\sigma\over 100\ {\rm km\ s}^{-1}}\right)^{-1.15} 
\label{eq:ndot1}\\
&\approx& 6.5\times 10^{-4} {\rm yr}^{-1} \left({\mh\over 10^6\msun}\right)^{-0.25}.
\label{eq:ndot2}
\end{eqnarray}
\end{mathletters}
\noindent
Merritt \& Ferrarese (2001a) find that $\log_{10}\mh/M_{bulge}$ is
distributed as a Gaussian with mean
$-2.91$ and dispersion $0.45$; the latter is consistent
with being due entirely to measurement errors in $\mh$.
Magorrian et al. (1998) find a mean mass-to-light ratio for their
galaxy sample of 
$\Upsilon_V \approx 4.9 h\left(L/10^{10}h^{-2}\lsun\right)^{0.18}\Upsilon_\odot$
($h\equiv H_0/80$ km s$^{-1}$ Mpc$^{-1}$).
Combining these relations with equation (\ref{eq:ndot2}) gives
\beq
\dot N \approx 2.2\times 10^{-4} {\rm yr}^{-1} h^{-0.25} 
\left({L\over 10^{10} h^{-2} \lsun}\right)^{-0.295}.
\label{eq:us}
\eeq
MT derived a similar relation 
(their ``toy model'', eq. 58) for consumption in a power-law nucleus.
Correcting for different assumed Hubble constants, their relation
is
\beq
\dot N \approx 2.6\times 10^{-5} {\rm yr}^{-1} h^{2/3} 
\left({L\over 10^{10} h^{-2} \lsun}\right)^{-0.22};
\label{eq:mt}
\eeq
the different scaling with $h$ results from their use of an
effective radius-luminosity relation in place of the $\mh-\sigma$
relation.
Our predicted event rate is a factor $\sim 7 (12)$ greater than theirs
at $L=10^{10}(10^{8})\lsun$;
these differences result primarily from the larger value
($0.006$ vs. $0.001$) assumed by MT for
$\langle \mh/M_{bulge}\rangle$,
and secondarily from our steeper dependence of $\dot N$ on $L$.

MT derived a total flaring rate for early type galaxies and bulges
by combining equation (\ref{eq:mt}) with the Ferguson \& Sandage (1991)
E + S0 luminosity function and assuming an equal contribution
from black holes in bulges.
They found a rate per unit volume of
$6.6\times 10^{-7}$ yr$^{-1}$ Mpc$^{-3}$ ($H_0=80$).
Comparing equations (\ref{eq:mt}) and (\ref{eq:us}),
we conclude that the downward revision in black hole masses
implies roughly an order of magnitude increase in the total event rate,
to $\sim 10^{-5}$ yr$^{-1}$ Mpc$^{-3}$.

The faintest systems in which there is solid kinematical evidence
for nuclear black holes are M32 and the bulge of the Milky Way
($L\approx 10^9\lsun, \mh\approx 10^{6.5}\msun$).
However there is compelling circumstantial evidence for supermassive
black holes in fainter systems (e.g. Filippenko \& Ho 2003) and less compelling
evidence for intermediate mass black holes
(IMBHs) in starburst galaxies and star clusters
(van der Marel 2003 and references therein).
Here we consider the consequences for the overall tidal flaring
rate if nuclear black holes exist in galaxies fainter than M32.
The galaxies in question are the dwarf ellipticals, spheroidal systems
fainter than $M_V\approx -19$ (\cite{FB94}).
In spite of their distinct name, dE galaxies have properties
that are a smooth continuation to lower luminosities of the properties
of bright ellipticals (\cite{HB97,GG03}).
The dEs are the most numerous type of galaxy in the universe;
in rich galaxy clusters their numbers appear to
diverge at low luminosities as $N(L)\sim L^{-1}$ 
(\cite{FS91}).
If dE galaxies contain nuclear black holes, they would
dominate the total tidal flaring rate due both to their numbers and
to their high individual event rates.

Rather than assume that every dE contains a nuclear black hole,
we make the more conservative assumption that only the
nucleated dEs (dEn's) contain black holes.
Most of the dEn's are too distant for their central
luminosity profiles to be resolved (e.g. Stiavelli et al. 2001);
one exception from the Local Group is
NGC 205, in which the deprojected density is observed to
increase as $\sim r^{-2}$ inward of 
$\sim 1$ pc (L. Ferrarese, private communication).
This is similar to what is seen in the other,
nucleated spheroidal systems in the Local Group
with comparable luminosities, namely M32 and
the bulges of M31 and M33 (\cite{Lauer98}) 
and the bulge of the Milky Way (\cite{Genzel03}).
We assume that all dEn's have nuclei with a similar structure
and that dEn's contain nuclear black holes with the same
ratio of black hole mass to total luminosity that is characteristic
of brighter galaxies.
We can then apply our scaling relations,
equations (\ref{eq:ndot1},b,\ref{eq:us}), to dEn galaxies.
We note in passing that the luminosity profiles of the dEns 
-- a steep nucleus superposed on a shallower 
background profile --
is just what is predicted by ``adiabatic growth'' models
for black holes (\cite{Peebles72,Young80}),
although there are other possible explanations for the origin
of the nuclei (e.g. Freeman 1993).

Van den Bergh (1986) plots the fraction of dEs that are nucleated
in a sample of galaxy clusters observed by Binggeli, Sandage
\& Tammann (1985).
He finds a roughly linear relation between the nucleated fraction
$F_n$ and absolute magnitude:
\beq
F_n\approx -0.2\left(M_V+13\right),\ \ \ \ -18\lap M_V\lap -13.
\label{eq:fn}
\eeq
The nucleated fraction is unity in dEs brighter than $M_V\approx -18$
and negligible in galaxies fainter than $M_V\approx -13$.
Trentham \& Tully (2002) derive luminosity functions for the dwarf
galaxy populations at the centers of
six galaxy clusters including the Virgo cluster.
They fit their data to a Schechter function,
\beq
N(M_R) dM_R = N_d\left({L\over L_d}\right)^{\alpha_d+1} e^{-L/L_d} dM_R
\label{eq:TT}
\eeq
with $M_R$ the $R$-band absolute magnitude.
The normalization factor $N_d$ has units of Mpc$^{-2}$ and gives
the surface density of dE galaxies at a distance of $200$ kpc from
the cluster center.
Trentham \& Tully find faint-end slopes of $-1.5\lap\alpha_d\lap -1$,
consistent with earlier determinations (e.g. Sandage, Binggeli \& Tamman
1985).

Table 2 gives the tidal flaring rate implied by equations
(\ref{eq:us}), (\ref{eq:fn}) and (\ref{eq:TT})
 for the centers of the six clusters analyzed by Trentham \& Tully (2002).
We give also an event rate for the center of the Coma cluster
based on the Secker \& Harris (1996) dE luminosity function.
The highest events rates, $\sim 0.1$ yr$^{-1}$ Mpc$^{-2}$,
are predicted for the centers of the Coma and Virgo clusters.

These predicted event rates could be substantially increased 
by including the contribution from the bulges of late-type spirals,
assuming the latter also contain IMBHs.
Bulge luminosity profiles are similar to those of dE galaxies
(\cite{MH01,Balcells03})
and often exhibit distinct nuclei (\cite{Carollo02}).
Balcells, Dominguez-Palmero \& Graham (2001) present resolved
nuclear density profiles in a sample of spiral bulges
observed with HST; the nuclei are well fit by power laws with 
$1.5\lap\gamma\lap 2.5$, similar to what is seen in the nuclei of the
Local Group dwarves.

The event rate due to dEn galaxies in the Virgo cluster
as a whole can be computed
using the determination by Ferguson \& Sandage (1989) 
of the spatial distribution of the dEn's.
They find a surface density $\Sigma(R)\approx \Sigma_0 e^{-R/R_0}$,
$R_0\approx 0.48$ Mpc, more centrally concentrated than the
distribution of non-nucleated dwarves.
Using the central density normalization of Trentham \& Tully (2002),
we find a total rate of tidal flaring due to dwarf
galaxies in Virgo of $\sim 0.16$ yr$^{-1}$.
Assuming Poission statistics,
the probability of detecting at least one event would be
$0.15$, $0.55$ and $0.80$ after $1$, $5$ and $10$ yr respectively
in the Virgo cluster alone.
While the spatial distribution of the dEn galaxies in the
Coma cluster has apparently not been determined,
we expect higher overall rates in Coma
than in Virgo due to its greater richness.

Some tidal flaring models (\cite{GO80,CLG90})
predict that single flares should persist for as long as several
months or years, 
and inspection of the light curves of the handful of candidate
X-ray events (\cite{KD02}) suggests decay times of this order.
Such long decay times would imply a non-trivial probability of observing an
ongoing disruption event {\it somewhere} in the Virgo or Coma cluster 
at any given time.
Non-detection of $X$-ray flares in these clusters
would constitute robust evidence that dE galaxies do not harbor 
IMBHs.

\clearpage
\section{Conclusions}

1. In most galaxies, the predicted rate of stellar tidal 
disruptions varies inversely with assumed black hole mass.
This is particularly true for galaxies with steep
central density profiles, which dominate the overall
event rate.

2. The downward revision in black hole masses that followed
the discovery of the $\mh-\sigma$ relation implies 
a total flaring rate per unit volume that is about an order 
of magnitude higher than in earlier studies.

3. An accurate analytic expression (equation \ref{eq:fapp})
can be derived that gives the tidal flaring rate as a function
of black hole mass and stellar velocity dispersion in
galaxies with $\rho\propto r^{-2}$ nuclei. 

4. If black holes are present in nucleated spheroids fainter than
$M_V\approx -19$, the tidal disruption rate due to dwarf galaxies in
the Virgo cluster would be of order $0.2$ yr$^{-1}$.
Non-detection of flares after a few years of monitoring
would argue against the existence of intermediate mass
black holes in dwarf galaxies.

We thank B. Binggeli, H. Cohn, A. Graham,
M. Milosavljevic, C. Pryor, and M. Stiavelli for useful discussions.
This work was supported by NSF grants
AST 00-71099 and AST02-0631
and by NASA grants NAG5-6037 and NAG5-9046.

\setcounter{table}{1}
\begin{deluxetable}{lr}
\tablecaption{Predicted Event Rates Due to Dwarf Galaxies \label{tbl-2}}
\tablewidth{0pt}
\tablehead{
\colhead{Name} & \colhead{Rate} \\
\colhead{ } & \colhead{(yr$^{-1}$ Mpc$^{-2}$)}
}
\startdata

Coma cluster & 0.093 \\
Virgo cluster & 0.11 \\
NGC 1407 group & 0.057 \\
Coma I & 0.032 \\
Leo group & 0.0063 \\
NGC 1023 group & 0.027 \\
Ursa Major group & 0.0079 \\

\enddata
\end{deluxetable}

\clearpage

\setcounter{table}{0}
\begin{deluxetable}{lrrrrrrrrrrrrr}
\tablefontsize{\scriptsize}
\tablecaption{Galaxy Sample\tablenotemark{a} \label{tbl-1}}
\tablewidth{0pt}
\tablehead{
\colhead{Name} & \colhead{Profile\tablenotemark{b}}   & \colhead{Distance}   &
\colhead{$\log_{10}(r_{b})$} &
\colhead{$\mu_b$}  & \colhead{$\alpha$} & \colhead{$\beta$} &
\colhead{$\Gamma$}  & \colhead{$\Upsilon_V $}   & 
\colhead{$\log_{10}(L_{V}/L_{\odot})$}  &
\colhead{$\log_{10}(M_{\bullet}/M_{\odot})$\tablenotemark{c}}  &
\colhead{$\log_{10}\dot{N}$\tablenotemark{d}} &
\colhead{$\log_{10}(M_{\bullet}/M_{\odot})$\tablenotemark{e}} &
\colhead{$\log_{10}\dot{N}$\tablenotemark{f}}\\
\colhead{ } & \colhead{ } & \colhead{(Mpc)} & \colhead{(pc)} &
\colhead{ } & \colhead{ } & \colhead{ } & \colhead{ } &
\colhead{($M_{\odot}/L_{\odot}$)} & \colhead{ } & \colhead{ } & 
\colhead{(yr$^{-1}$)} & \colhead{ }  & \colhead{(yr$^{-1}$)} 
}
\startdata
NGC221 & $\setminus$  &0.8 &-0.26 &11.77 &0.98 &1.36 &0.01 &2.27 &8.57  &6.38 &-3.79 &6.32 &-3.78\\
NGC224 & $\setminus$  &0.8  &0.11 &13.44 &4.72 &0.81 &0.12 &26.1 &9.86  &7.79 &-3.70 &6.13 &-3.56\\
NGC596 & $\setminus$  &21.2 &2.56 &18.03 &0.76 &1.97 &0.55 &4.16 &10.29 &8.69 &-5.01 &7.69 &-4.52 \\
NGC1023 & $\setminus$ &10.2 &1.96 &16.17 &4.72 &1.18 &0.78 &5.99 &9.99  &8.55 &-4.46 &8.17 &-4.19 \\
NGC1172 & $\setminus$ &29.8 &2.55 &18.61 &1.52 &1.64 &1.01 &2.57 &10.23 &8.42 &-4.75 &6.90 &-3.24\\
NGC1426 & $\setminus$ &21.5 &2.23 &17.53 &3.62 &1.35 &0.85 &4.91 &10.07 &8.54 &-4.87 &7.50 &-4.08\\
NGC3115 & $\setminus$ &8.4  &2.07 &16.17 &1.47 &1.43 &0.78 &7.14 &10.23 &8.61 &-4.19 &8.74 &-4.28\\
NGC3377 & $\setminus$ &9.9  &0.64 &12.85 &1.92 &1.33 &0.29 &2.88 &9.81  &7.79 &-4.16 &7.51 &-4.04\\
NGC3599 & $\setminus$ &20.3 &2.12 &17.58 &13.0 &1.66 &0.79 &2.09 &9.82  &7.91 &-5.24 &6.22 &-4.15\\
NGC3605 & $\setminus$ &20.3 &1.94 &17.25 &9.14 &1.26 &0.67 &4.05 &9.59  &8.10 &-5.17 &6.76 &-4.50\\
NGC4239 & $\setminus$ &15.3 &1.98 &18.37 &14.5 &0.96 &0.65 &3.37 &9.19  &7.49 &-5.57 &5.69 &-4.84\\
NGC4387 & $\setminus$ &15.3 &2.52 &18.89 &3.36 &1.59 &0.72 &5.34 &9.48  &7.99 &-5.13 &6.83 &-4.46\\
NGC4434 & $\setminus$ &15.3 &2.25 &18.21 &0.98 &1.78 &0.70 &4.73 &9.52  &7.97 &-4.81 &6.81 &-4.16\\
NGC4458 & $\setminus$ &15.3 &0.95 &14.49 &5.26 &1.43 &0.49 &4.00 &9.52  &7.90 &-4.58 &6.78 &-4.23\\
NGC4464 & $\setminus$ &15.3 &1.95 &17.35 &1.64 &1.68 &0.88 &4.82 &9.22  &7.69 &-4.21 &7.26 &-3.88\\
NGC4467 & $\setminus$ &15.3 &2.38 &19.98 &7.52 &2.13 &0.98 &6.27 &8.75  &7.32 &-4.48 &6.04 &-3.30 \\
NGC4478 & $\setminus$ &15.3 &1.10 &15.40 &3.32 &0.84 &0.43 &5.03 &9.79  &8.27 &-5.00 &7.34 &-4.64\\
NGC4551 & $\setminus$ &15.3 &2.46 &18.83 &2.94 &1.23 &0.80 &7.25 &9.57  &8.21 &-4.96 &7.11 &-4.19\\
NGC4564 & $\setminus$ &15.3 &1.59 &15.70 &0.25 &1.90 &0.05 &4.48 &9.91  &8.40 &-4.67 &7.71 &-4.27\\
NGC4570 & $\setminus$ &15.3 &2.32 &17.29 &3.72 &1.49 &0.85 &5.52 &9.95  &8.47 &-4.49 &8.01 &-4.14\\
NGC4621 & $\setminus$ &15.3 &2.34 &17.20 &0.19 &1.71 &0.50 &6.73 &10.44 &8.45 &-4.04 &8.49 &-4.07\\
NGC4697 & $\setminus$ &10.5 &2.12 &16.93 &24.9 &1.04 &0.74 &6.78 &10.34 &8.95 &-5.03 &7.73 &-4.25 \\
NGC4742 & $\setminus$ &12.5 &1.93 &16.69 &48.6 &1.99 &1.09 &1.76 &9.62  &7.65 &-3.80 &6.85 &-2.90\\
NGC5845 & $\setminus$ &28.2 &2.49 &17.52 &1.27 &2.74 &0.51 &6.69 &9.88  &8.48 &-4.56 &8.70 &-4.66 \\
NGC7332 & $\setminus$ &20.3 &1.88 &15.72 &4.25 &1.34 &0.90 &1.56 &9.90  &7.87 &-4.33 &7.21 &-3.78 \\
\\
A2052   & $\cap$ &132.0 &2.43 &18.36 &8.02 &0.75 &0.20 &12.80 &11.00 &9.88  &-5.65 &8.62 &-4.90\\
NGC720  & $\cap$ &22.6  &2.55 &17.50 &2.32 &1.66 &0.06 &8.15  &10.58 &9.27  &-5.47 &8.51 &-5.50\\
NGC1399 &$\cap$  &17.9  &2.43 &17.06 &1.50 &1.68 &0.07 &12.73 &10.62 &9.72  &-5.22 &9.08 &-5.09\\
NGC1600 & $\cap$ &50.2  &2.88 &18.38 &1.98 &1.50 &0.08 &14.30 &11.01 &10.06 &-5.71 &9.11 &-5.62\\
NGC3379 & $\cap$ &9.9   &1.92 &16.10 &1.59 &1.43 &0.18 &6.87  &10.15 &8.59  &-4.90 &8.30 &-4.85\\
NGC4168 & $\cap$ &36.4  &2.65 &18.33 &0.95 &1.50 &0.14 &7.54  &10.64 &9.08  &-5.59 &7.89 &-5.47\\
NGC4365 & $\cap$ &22.0  &2.25 &16.77 &2.06 &1.27 &0.15 &8.40  &10.76 &9.46  &-5.29 &8.57 &-5.14\\
NGC4472 & $\cap$ &15.3  &2.25 &16.66 &2.08 &1.17 &0.04 &9.20  &10.96 &9.42  &-5.15 &8.79 &-5.05\\
NGC4486 & $\cap$ &15.3  &2.75 &17.86 &2.82 &1.39 &0.25 &17.70 &10.88 &9.56  &-5.35 &9.16 &-5.28\\
NGC4486b& $\cap$ &15.3  &1.13 &14.92 &2.78 &1.33 &0.14 &9.85  &8.96  &8.96  &-4.84 &8.17 &-4.43\\
NGC4636 & $\cap$ &15.3  &2.38 &17.72 &1.64 &1.33 &0.13 &10.40 &10.60 &8.36  &-5.35 &8.07 &-5.37\\
NGC4649 & $\cap$ &15.3  &2.42 &17.17 &2.00 &1.30 &0.15 &16.20 &10.79 &9.59  &-5.19 &9.19 &-5.12\\
NGC4874 & $\cap$ &93.3  &3.08 &19.18 &2.33 &1.37 &0.13 &15.00 &11.35 &10.32 &-6.02 &8.77 &-5.91\\
NGC4889 & $\cap$ &93.3  &2.88 &18.01 &2.61 &1.35 &0.05 &11.20 &11.28 &10.43 &-5.81 &9.20 &-5.69\\
NGC5813 & $\cap$ &28.3  &2.04 &16.42 &2.15 &1.33 &0.08 &7.10  &10.66 &9.29  &-5.24 &8.27 &-5.10\\
NGC6166 & $\cap$ &112.5 &3.08 &19.35 &3.32 &0.99 &0.08 &15.60 &11.32 &10.47 &-6.16 &8.84 &-6.14\\
\\
NGC524  & $\cap$      &23.1  &1.55 &16.02 &1.29 &1.00  &0.00 &14.30 &10.54 &9.47  &-  &8.62  &-\\
NGC1316 & $\cap$      &17.9  &1.55 &14.43 &1.16 &1.00  &0.00 &2.56  &11.06 &9.25  &-  &8.36  &-\\
NGC1400 & $\cap$      &21.5  &1.54 &15.41 &1.39 &1.32  &0.00 &10.70 &10.36 &9.16  &-  &8.62  &-\\
NGC1700 & $\setminus$ &35.5  &1.19 &13.95 &0.90 &1.30  &0.00 &4.00  &10.59 &8.97  &-  &8.38  &-\\
NGC2636 & $\setminus$ &33.5  &1.17 &15.68 &1.84 &1.14  &0.04 &2.97  &9.47  &7.72  &-  &6.52  &-\\
NGC2832 & $\cap$      &90.2  &2.60 &17.45 &1.84 & 1.40 &0.02 &10.90 &11.11 &10.06 &-  &9.05  &-\\
NGC2841 & $\setminus$ &13.2  &0.92 &14.55 &0.93 & 1.02 &0.01 &8.98  &9.88  &8.62  &-  &8.23  &-\\
NGC3608 & $\cap$      &20.3  &1.44 &15.45 &1.05 &1.33  &0.00 &7.04  &10.27 &8.39  &-  &8.01  &-\\
NGC4552 & $\cap$      &15.3  &1.68 &15.41 &1.48 &1.30  &0.00 &7.66  &10.35 &8.67  &-  &8.62  &-\\
NGC7768 & $\cap$      &103.1 &2.30 &16.99 &1.92 & 1.21 &0.00 &9.51  &11.10 &9.93  &-  &8.82  &-\\

 \enddata

\tablenotetext{a}{All parameters except for black hole mass 
and consumption rate are taken from Faber et al. (1997)
and assume $H_0=80$ km s$^{-1}$ Mpc$^{-1}$.}

\tablenotetext{b}{Profile class: $\cap$=core galaxy; $\setminus$=power law galaxy}

\tablenotetext{c}{Black hole mass from Magorrian et al. (1998).}

\tablenotetext{d}{Consumption rate based on the Magorrian et al. (1998)
black hole mass.}

\tablenotetext{e}{Black hole mass from the $\mh-\sigma$
relation, equation (\ref{eq:ms}).}

\tablenotetext{f}{Consumption rate based on the $\mh-\sigma$ black
hole mass.}

\end{deluxetable}


\begin{thebibliography}{}

\bibitem[Balcells et al. 2001]{Balcells01}
	Balcells, M., Dominguez-Palmero, L. \&
	Graham, A. 2001,
	in ASP Conf. Ser. 249, The Central Kiloparsec of Starbursts
	and AGN,
	ed. J. H. Knapen et al.
        (San Francisco: ASP), 167

\bibitem[Balcells et al. 2003]{Balcells03}
	Balcells, M., Graham, A. W., Dominguez-Palmero, L. \&
	Peletier, R. 2003,
	ApJ, 582, L79	

\bibitem[Binggeli, Sandage \& Tamman 1985]{BST85}
	Binggeli, B., Sandage, A., \& Tamman, G. A. 1985,
	AJ, 90, 1681

\bibitem[Cannizzo, Lee, \& Goodman 1990]{CLG90}
	Cannizzo, J. K., Lee, H. M., \& Goodman, J. 1990,
	ApJ, 351, 38

\bibitem[Carollo et al. 2002]{Carollo02}
	Carollo, C. M., Stiavelli, M., Seigar, M., de Zeeuw, P. T.
	\& Dejonghe, H. 2002,
	ApJ, 123, 159

\bibitem[Cohn \& Kulsrud 1978]{CK78} 
	Cohn, H., \& Kulsrud, R. M. 1978, 
	ApJ, 226, 1087 (CK)

\bibitem[Donley et al. 2002]{Donley02}
	Donley, J. L., Brandt, W. N., Eracleous, M., \& Boller, Th. 2002,
	AJ, 124, 1308

\bibitem[Erwin, Graham \& Caon 2002]{Erwin02}
	Erwin, P., Graham, A. W. \& Caon, N. 2003,
	preprint
	(astro-ph/0212335)

\bibitem[Evans \& Kochanek 1989]{EK89}
	Evans, C. R. \& Kochanek, C. S. 1989,
	ApJ, 346, L13

\bibitem[Faber et al. 1997]{Faber97} 
	Faber, S. M. et al. 1997, 
	AJ, 114, 1771

\bibitem[Ferguson \& Binggeli 1994]{FB94}
	Ferguson, H. C., \& Binggeli, B. 1994,
	Astron. Ap. Rev. 6, 67

\bibitem[Ferguson \& Sandage 1989]{FS89}
	Ferguson, H. C., \& Sandage, A. 1989,
	AJ, 346, L53

\bibitem[Ferguson \& Sandage 1991]{FS91}
	Ferguson, H. C., \& Sandage, A. 1991,
	AJ, 101, 765

\bibitem[Ferrarese \& Merritt 2000]{FM00}
	Ferrarese, L., \& Merritt, D. 2000,
	ApJ, 539, L9

\bibitem[Filippenko \& Ho 2003]{FH03}
	Filippenko, A., \& Ho, L. C. 2003,
	preprint
	(astro-ph/0303429)
 
\bibitem[Frank \& Rees 1976]{FR76} 
	Frank, J., \& Rees, M. J. 1976, 
	MNRAS, 176, 633 (FR)

\bibitem[Freeman 1993]{Freeman93}
	Freeman, K. 1993,
	in ASP Conf. Ser. 48, The Globular Cluster-Galaxy Connection,
	ed. G. H. Smith \& J. P. Brodie
        (San Francisco: ASP), 608

\bibitem[Gebhardt et al. 1996]{Gebhardt96}
	Gebhardt, K. et al. 1996,
	AJ, 112, 105

\bibitem[Gebhardt et al. 2000]{Gebhardt00}
	Gebhardt, K. et al. 2000,
	ApJ, 539, L13

\bibitem[Genzel et al. 2003]{Genzel03}
	Genzel, R. et al. 2003,
	preprint
	(astro-ph/0305423)

\bibitem[Gezari et al. 2003]{Gezari03}
	Gezari, S., Halpern, J. P., Komossa, S., Grupe, D., \& Leighly, K. M.
	2003,
	preprint
	(astro-ph/0304063)

\bibitem[Graham \& Guzman 2003]{GG03}
	Graham, A. W. \& Guzman, R. 2003,
	preprint
	(astro-ph/0303391)

\bibitem[Gurzadyan \& Ozernoy 1980]{GO80}
	Gurzadyan, V. G. \& Ozernoy, L. M. 1980,
	AA, 86, 315

\bibitem[Jerjen \& Binggeli 1997]{HB97}
	Jerjen, H. \& Binggeli, B. 1997,
	in The Nature of Elliptical Galaxies,
	ASP Conf. Ser. Vol. 116,
	ed. M. Arnaboldi, G. S. Da Costa, \& P. Saha
        (San Francisco: ASP), 239

\bibitem[Hills 1975]{Hills75}
	Hills, J. G. 1975,
	Nature, 254, 295

\bibitem[Kim, Park \& Lee 1999]{Kim99}
	Kim, S. S., Park, M.-G., \& Lee, H. M. 1999,
	ApJ, 519, 647

\bibitem[Komossa 2002]{Komossa02}
	Komossa, S. 2002,
	Rev. Mod. Astron. 15, 27 

\bibitem[Komossa \& Dahlem 2002]{KD02}
	Komossa, S., \& Dahlem, M. 2002,
	preprint
	(astro-ph/0106422)

\bibitem[Kormendy \& Gebhardt 2001]{KG01}
	Kormendy, J. \& Gebhardt, K. 2001,
	AIP Conf. Proc. Vol. 586,
        ed. J. C. Wheeler \& H. Martel,
	363
 
\bibitem[Lauer et al. 1998]{Lauer98}
	Lauer, T. et al. 1998,
	AJ, 116, 2263

\bibitem[Lidskii \& Ozernoi 1979]{LO79}
	Lidskii, V. V. \& Ozernoi, L. M. 1979,
	Pis'ma Astron. Zh. 5, 28

\bibitem[Magorrian et al. 1998]{Magorrian98} 
	Magorrian, J., et al. 1998, 
	AJ, 115, 2285

\bibitem[Magorrian \& Tremaine 1999]{MT99} 
	Magorrian, J., \& Tremaine, S. 1999, 
	MNRAS, 309, 447 (MT)

\bibitem[McLure \& Dunlop 2002]{MD02}
	McLure, R. J. \& Dunlop, J. S. 2002,
	MNRAS, 331, 795

\bibitem[Merritt 2003]{Merritt03}
	Merritt, D. 2003,
	preprint
	(astro-ph/0301257) 

\bibitem[Merritt \& Ferrarese 2001a]{MF01a}
	Merritt, D., \& Ferrarse, L. 2001a,
	MNRAS, 320, L30

\bibitem[Merritt \& Ferrarese 2001]{MF01b} 
	Merritt, D., \& Ferrarese, L. 2001b, 
	in ASP Conf. Ser. 249, 
	The Central Kiloparsec of Starbursts and AGN: 
	The La Palma Connection, 
	ed. H. Knapen, J. E. Beckman, I. Shlosman, \& 
	T. J. Mahoney. (San Francisco: ASP), 335

\bibitem[Merritt \& Fridman 1996]{MF96}
	Merritt, D. \& Fridman, T. 1996,
	in ASP Conf. Ser. 86,
	Fresh Views of Elliptical Galaxies,
	ed. A. Buzzoni, A. Renzini, \& A. Serrano
	(San Francisco: ASP), 13

\bibitem[Milosavljevic \& Merritt 2003]{MM03}
	Milosavljevic, M. \& Merritt, D. 2003,
	preprint
	(astro-ph/0212459)

\bibitem[M\"ollenhoff \& Heidt 2001]{MH01}
	M\"ollenhoff, C. \& Heidt, J. 2001,
	A\&A, 368, 16

\bibitem[Peebles 1972]{Peebles72}
	Peebles, P. J. E. 1972,
	ApJ, 178, 371

\bibitem[Rees 1988]{Rees88}
	Rees, M. J. 1988,
	Nature, 333, 523

\bibitem[Rees 1998]{Rees98}
	Rees, M. J. 1998,
	in Black Holes and Relativistic Stars,
	ed. R. M. Wald (Chicago: University of Chicago Press),
	79

\bibitem[Richstone et al. 1998]{Richstone98}
	Richstone, D. O. et al. 1998,
	Nature, 395, A14

\bibitem[Sandage, Binggeli, \& Tamman 1985]{SBT85}
	Sandage, A., Binggeli, B., \& Tamman, G. A. 1985,
	AJ, 90, 1759

\bibitem[Secker \& Harris 1996]{SH96}
	Secker, J. \& Harris, W. E. 1996,
	ApJ, 469, 623

\bibitem[Spitzer \& Hart 1971]{SH71}
	Spitzer, L. \& Hart, M. H. 1971,
	ApJ, 164, 399

\bibitem[Spitzer \& Harm 1958]{SH58}
	Spitzer, L. \& Harm, R. 1958,
	ApJ, 127, 544

\bibitem[Stiavelli et al. 2001]{Stiavelli01}
	Stiavelli, M., Miller, B. W., Ferguson, H. C.,
	Mack, J., Whitmore, B. C., \& Lotz, J. M. 2001,
	AJ, 121, 1385

\bibitem[Syer \& Ulmer 1999]{SU99} 
	Syer D., \& Ulmer A. 1999, 
	MNRAS, 306, 35 (SU)

\bibitem[Trentham \& Tully 2002]{TT02}
	Trentham, N. \& Tully, R. B. 2002,
	MNRAS, 335, 712

\bibitem[Ulmer 1999]{Ulmer99}
	Ulmer, A. 1999,
	ApJ,514, 180

\bibitem[van den Bergh 1986]{vdb86}
	van den Bergh, S. 1986,
	AJ, 91, 271

\bibitem[van der Marel 2003]{vdm03}
	van der Marel, R. 2003,
	preprint
	(astro-ph/0302101)

\bibitem[Wandel 1999]{Wandel99}
	Wandel, A. 1999,
	ApJ, 519, 39

\bibitem[Young 1980]{Young80}
	Young, P. 1980,
	ApJ, 242, 1232

\end{thebibliography}
\end{document}